\documentclass[letterpaper,12pt]{article}
\usepackage{amssymb,enumerate,float,amsmath}
\usepackage[title]{appendix}

\usepackage{ifpdf}
\ifpdf
\usepackage[pdftex]{graphicx}
\usepackage[pdftex,unicode,implicit]{hyperref}

\hypersetup{
	pdftitle     = {}, 
	pdfkeywords  = {},
	pdfauthor    = {},
	pdfcreator   = {pdf\LaTeXe\ with package \flqq hyperref\frqq},
	pdfproducer  = {pdf\LaTeXe\ with package \flqq hyperref\frqq},
	pdfpagemode  = UseNone,  
	pdffitwindow = true,  
	unicode      = true,
	plainpages   = true,
	colorlinks   = true,  
	citecolor    = black,  
	urlcolor     = blue, 
	linkcolor    = black
}

\else

\usepackage[dvips]{graphicx}

\fi

\makeatletter
\@addtoreset{equation}{section}
\makeatother

\begin{document}
	
\thispagestyle{empty}
	
\begin{center}
{\bf \LARGE BRST symmetry and unitarity of the Ho\v{r}ava theory}
\vspace*{15mm}
		
{\large Jorge Bellor\'{\i}n}$^{1,a}$,
{\large Claudio B\'orquez$^{1,b}$}
{\large and Byron Droguett}$^{2,c}$
\vspace{3ex}

$^1${\it Department of Physics, Universidad de Antofagasta, 1240000 Antofagasta, Chile.}

$^2${\it Departamento de Ciencias B\'asicas, Facultad de Ciencias, Universidad Santo Tom\'as, Sede Arica 1000000, Chile.}
\vspace{3ex}

$^a${\tt jorge.bellorin@uantof.cl,} \hspace{1em}
$^b${\tt cl.borquezg@gmail.com,} \hspace{1em}
$^c${\tt byrondroguett@santotomas.cl}
		
\vspace*{15mm}
{\bf Abstract}
\begin{quotation}{ \small \noindent
We present an analysis on the BRST symmetry transformations of the Ho\v{r}ava theory under the BFV quantization, both in the nonprojectable and projectable cases. We obtain that the BRST transformations are intimately related to a particular spatial diffeomorphism along one of the ghost vector fields. We show explicitly the invariance of the quantum action and the nilpotence of the BRST transformations, using this diffeomorphism largely. The BRST symmetry is verified in the whole phase space, inside and outside the constrained surface. When restricted to the constrained surface, the BRST transformations are completely local. The consistency of the BRST symmetry is a fundamental feature of a quantum field theory, specially for renormalization. The BFV quantization is independent of the chosen gauge-fixing condition. This allows us to study the unitarity of the quantum Ho\v{r}ava theory, covering the gauge required for renormalization. We prove that, if the solution for  the lapse function exists, then the unitarity of the theory is established.
}
\end{quotation}
\end{center}

\newpage

\section{Introduction}
There have been important advances in the quantization of the Ho\v{r}ava theory \cite{Horava:2009uw}, which is a proposal of a theory of quantum gravity aimed to solve the problem of the (perturbative) nonrenormalizability of general relativity. The theory introduces a foliation of spacelike hypersurfaces along a timelike direction, such that the foliation has an absolute physical meaning. The group of diffeomorphisms that preserve the given foliation is the gauge group of the theory. In this way, spatial derivatives are increased without increasing the order in time derivatives. This is the core for the renormalizability. There are two different versions of the theory that are compatible with this gauge symmetry: the nonprojectable and the projectable cases. Their definitions are given in the next section. The nonprojectable version has field equations that, at the large-distance limit, are closer to the Einstein equations than the ones of the projectable case. From this point of view, the nonprojectable version acquires a central relevance in the context of the Ho\v{r}ava theory.

A significant advance in the quantization is the proof of the complete renormalization of the projectable case presented in Ref.~\cite{Barvinsky:2015kil}; see also Ref.~\cite{Barvinsky:2017zlx}. The projectable version is a theory with only first-class constraints, which implies that, in terms of the Lagrangian formulation, it can be quantized as the class of gauge theories as Yang-Mills theory or perturbative general relativity. A remarkable feature of the renormalization showed in \cite{Barvinsky:2015kil} is the introduction of a nonlocal gauge-fixing condition; hence a nonlocal quantum Lagrangian. On the other hand, standard renormalizability requires a local Lagrangian. The reconciliation between these two aspects is achieved in the Hamiltonian formulation: the canonical Lagrangian with the required gauge-fixing condition is indeed local. This was first observed in Ref.~\cite{Barvinsky:2015kil}. In Ref.~\cite{Bellorin:2021udn} we performed the complete quantization of the projectable case in the Hamiltonian formalism using the Batalin-Fradkin-Vilkovisky (BFV) scheme of quantization \cite{Fradkin:1975cq,Batalin:1977pb,Fradkin:1977xi}. The starting point to motivate this analysis is that the gauge-fixing condition required for renormalization is a noncanonical condition. The BFV quantization is suitable for such a gauge condition since it was precisely designed for relativistic gauges that are not canonical, like the Lorentz gauge for Yang-Mills theory, when the problem of the unitarity of those theories is focused. In \cite{Bellorin:2021udn}, we obtained the local Hamiltonian, and the integration on momenta led to the same quantum Lagrangian of Ref.~\cite{Barvinsky:2015kil} with the nonlocal gauge-fixing condition. A further advance in the quantization of the projectable theory is the proof that the $2+1$ theory is asymptotically free \cite{Barvinsky:2017kob}, among other developments that can be found in the literature.

The quantization of the nonprojectable case presents a fundamental difference with respect to the projectable case: the presence of second-class constraints. Indeed, the proof of renormalization of the nonprojectable case is still pending. We have studied the quantization of the nonprojectable version using also the BFV formalism \cite{Bellorin:2021udn,Bellorin:2021tkk,Bellorin:2022qeu}. In this scheme the same noncanonical gauge-fixing condition of the projectable case \cite{Barvinsky:2015kil}, together with the second-class constraints can be incorporated. The main point is that the second-class constraints affect the measure of the path integral \cite{Senjanovic:1976br}. The analysis of Ref.~\cite{Bellorin:2019gsc} shows that the measure of the second-class constraints changes the behavior of some propagators, pointing to regular propagators, as required for renormalization \cite{Barvinsky:2015kil}. We obtained a further advance using the BFV formalism: in Ref.~\cite{Bellorin:2022qeu} we found that, although the quantum measure leads still to nonregular propagators for some fields, the divergences of loops formed with these fields cancel between them completely. The rest are loops formed with regular propagators. These results are encouraging to continue on the program of the quantization of the nonprojectable case under the BFV formalism.

A fundamental aspect of the quantization of a field theory with gauge symmetry is the Becchi-Rouet-Stora-Tyutin (BRST) symmetry, specially to get the final renormalization. The BFV quantization is intimately linked to this class of symmetry \cite{Batalin:1977pb}. The Hamiltonian is presented as a gauge-fixed Hamiltonian, where the gauge fixing is done by operating with the BRST charge. The main theorem developed in the original papers of the BFV formalism ensures that the path integral is independent of the chosen gauge-fixing function. The BRST charge is formally defined using  the first-class constraints and their gauge algebra for a given theory. Its action on the fields is defined by a canonical transformation: with Poisson brackets for the case of a theory with first-class constraints only, and with Dirac brackets when second-class constraints are present. Thus, in the case of the nonprojectable theory the whole procedure applies using Dirac brackets.

Our first aim in this paper is to develop a detailed study of the BRST symmetry in both versions of the Ho\v{r}ava theory. We obtain explicit expressions for these transformations and study how the invariance of the action is achieved. We show this explicitly, beyond the formal definitions of the general BFV quantization. Since we pursue to make a general study, we develop the BRST transformations on nonperturbative variables and in the whole phase space, including the subset where the second-class constraints are not satisfied (the exception is the transformation of the gauge-fixing terms, which are defined in terms of perturbative variables). We find the interesting result that most of the BRST transformations can be modeled in terms of a spatial diffeomorphism along one of the ghost vector fields. We take this result as the basis to operate the BRST symmetry on the quantum action. Moreover, we show explicitly the nilpotence of the BRST transformations, where the diffeomorphism along the ghost field plays a prominent role. For a possible application to renormalization, it is important to elucidate the local or nonlocal character of the BRST transformations. We discuss how this is related to the second-class constraints.

Our second objective is to take advantage of the well-posed BFV quantization to study the unitarity of the Ho\v{r}ava theory, which is a fundamental feature of the $S$ matrix. We remark that the BFV formalism was developed precisely to show the unitarity of relativistic theories (Yang-Mills theory, general relativity) when relativistic gauges are used. The unitarity is achieved by making use of the capacity of the formalism to change the gauge condition. In this paper we show that, once the BFV quantization has been completed under an appropriate gauge condition, and assuming that the solution of a second-class constraint exists, the integration on all unphysical variables leads to a canonical path integral with the correct degrees of freedom on the integration and with measure 1. We give evidence for the existence of the solution of the second-class constraint under certain limits.


\section{BFV quantization of the nonprojectable case}
Ho\v{r}ava's gravitational theory is formulated using the Arnowitt-Deser-Misner (ADM) variables of the given foliation, $N(t,\vec{x})$, $N^i(t,\vec{x})$, and $g_{ij}(t,\vec{x})$. The foliation-preserving diffeomorphisms are defined by $\delta t=f(t)$, $\delta x^{i}=\zeta^{i}(t,\vec{x})$. Their action of the ADM variables is
\begin{eqnarray}
	\delta N &=& 
	\zeta^{k} \partial_{k} N + f \dot{N} + \dot{f}N \,, 
	\label{deltadiffN}
	\\ 
	\delta N_{i} &=& 
	\zeta^{k} \partial_{k} N_{i} + N_{k} \partial_{i} \zeta^{k} + \dot{\zeta}^{j} g_{ij} + f \dot{N}_{i} + \dot{f}N_{i} \,,
	\\
	\delta g_{ij} &=& \zeta^{k} \partial_{k} g_{ij} + 2g_{k(i}\partial_{j)} \zeta^{k} + f \dot{g}_{ij} \,.
\end{eqnarray}
Among these transformations, only the spatial diffeomorphisms ($f=0$, $\zeta^i(t,\vec{x})$ arbitrary) constitute a gauge symmetry in the strict sense. The nonprojectable case is defined by the condition of the lapse function $N(t,\vec{x})$ that can be a function of time and the space, whereas in the projectable case it is restricted to be a function only of time (and, consequently, it is an spurious degree of freedom). The nonprojectable case exhibits a dynamics closer to the one of general relativity. In this section, we study the nonprojectable case.

Given the foliation, the action of the nonprojectable theory is \cite{Horava:2009uw,Blas:2009qj}
\begin{equation}
 S = \int dt d^2x \sqrt{g} N \left( K_{ij} K^{ij} - \lambda K^2 - \mathcal{V} \right) \,.
 \label{classicalaction}
\end{equation}
The kinetic terms of the Lagrangian are defined in terms of the extrinsic curvature tensor,
\begin{equation}
 K_{ij} = \frac{1}{2N} \left( \dot{g}_{ij} - 2 \nabla_{(i} N_{j)} \right) \,,
 \qquad
 K = g^{ij} K_{ij} \,.
\end{equation}
$\mathcal{V}$ is called the potential. It contains the higher-order spatial derivatives, whose order is identified with the parameter $z$, such that the higher-order terms are of $2z$ order in spatial derivatives. In this paper it is sufficient to consider the $(2+1)$-dimensional theory, since the $3+1$ case is conceptually equal. The $2+1$ theory requires a potential of $z=2$ order for power-counting renormalizability. The complete potential is \cite{Sotiriou:2011dr}
\begin{eqnarray}
\mathcal{V} &=&
-\beta R-\alpha a^{2}+\alpha_{1}R^{2}+\alpha_{2}a^{4}+\alpha_{3}R a^{2}+\alpha_{4}a^{2}\nabla_{i}a^{i}
\nonumber \\ &&
+\alpha_{5} R\nabla_{i}a^{i} 
+\alpha_{6}\nabla^{i}a^{j}\nabla_{i}a_{j}+ \alpha_{7}(\nabla_{i}a^{i})^{2} \,,
\label{potencial}
\end{eqnarray}
where 
\begin{equation}
a_i = \frac{\partial_i N}{N} \,,
\quad
a^2 \equiv a_k a^k \,.
\end{equation} 
$\nabla_i$ and $R$ are the covariant derivative and the Ricci scalar of the spatial metric $g_{ij}$. The coupling constants of the theory are $\lambda$, $\beta$, $\alpha$, $\alpha_1$,...,$\alpha_7$. We use shorthand for frequent combinations of these coupling constants,
\begin{equation}
 \sigma = \frac{1-\lambda}{1-2\lambda} \,,
 \quad
 \bar{\sigma} = \frac{\lambda}{1-2\lambda} \,,
 \quad
 \alpha_{67} = \alpha_6 + \alpha_7 \,.
\end{equation} 

In the Hamiltonian formulation \cite{Kluson:2010nf,Donnelly:2011df,Bellorin:2011ff} the canonically conjugate pairs are $(g_{ij},\pi^{ij})$ and $(N,P_N)$. After the Legendre transformation, the primary classical Hamiltonian results in
\begin{equation}
	H_0 =
	\int d^{2}x \mathcal{H}_0 \,,
	\quad
	\mathcal{H}_0 \equiv	
	\sqrt{g} N \left( \frac{\pi^{ij}\pi_{ij}}{g} 
	+ \bar{\sigma} \frac{\pi^{2}}{g} + \mathcal{V} \right) \,.
	\label{H0}
\end{equation}
The constraint that corresponds to the involutive functions under Dirac brackets in the BFV quantization is  
\begin{equation}
	\mathcal{H}_i \equiv
	- 2 g_{ij} \nabla_k \pi^{kj} = 0 \,.
	\label{momentumconts}
\end{equation}
As it is well known from general relativity, this is the generator of the spatial diffeomorphisms on the canonical pair $( g_{ij} , \pi^{ij} )$. The second-class constraints are 
\begin{eqnarray}
	\theta_{1} &\equiv&  
	\mathcal{H}_0 
	+ \sqrt{g} \Big( 
	  2\alpha\nabla_{i}(Na^{i})
	- 4\alpha_{2}\nabla_{i}(Na^{2}a^{i})
	- 2\alpha_{3}\nabla_{i}(NRa^{i})
	+ 2\alpha_{6}\nabla^{i}\nabla^{j}(N\nabla_{j}a_{i})
	\nonumber\\ &&
	+ \alpha_{4} \left( \nabla^{2}(Na^{2}) 
	-2\nabla_{i}( N a^{i} \nabla_{j} a^{j} ) \right)
	+ \alpha_{5}\nabla^{2}(NR)	
	+ 2\alpha_{7}\nabla^{2}(N\nabla_{i}a^{i})
    \Big) 
    = 0 \,,
	\nonumber \\ &&
	\label{theta1}
	\\ 
	\theta_{2} &\equiv& P_{N}=0 \,.
    \label{theta2}
\end{eqnarray}
An important feature of the nonprojectable theory is that the primary Hamiltonian (\ref{H0}) can be written as the integral of one of the second-class constraints \cite{Bellorin:2019zho},
\begin{equation}
	H_{0} = \int d^{2}x \,\theta_{1} \,.
	\label{primaryhamiltonian}
\end{equation}
This holds because the terms in $\theta_1$ that are not contained in $\mathcal{H}_0$ are total derivatives that vanish after spatial integration.\footnote{In $d=3$ spatial dimensions, there are nonzero boundary terms in (\ref{primaryhamiltonian}). They ensure the differentiability of the functional, similar to the Hamiltonian formulation of general relativity (see Refs.~\cite{Donnelly:2011df,Bellorin:2011ff}).} The phase space of the classical theory is spanned by $\{g_{ij},\pi^{ij},N,P_N\}$. These are eight functional degrees of freedom. The theory has the constraints $\mathcal{H}_i$, $\theta_1$, and $\theta_2$, which sum up four conditions. The gauge symmetry of the spatial diffeomorphism eliminate two more degrees, leaving a physical phase space of dimension two. This is the physical phase space of a scalar mode.\footnote{In three spatial dimensions the field variables $\{g_{ij},\pi^{ij},N,P_N\}$ sum $14$, the constraints $\mathcal{H}_i$, $\theta_1$, and $\theta_2$ give five and the spatial diffeomorphisms three. The result is a physical phase space of dimension six: two tensorial modes and one extra scalar mode.}

We present a summary of the BFV quantization of the nonprojectable Ho\v{r}ava theory using a specific gauge-fixing condition. Further details about the procedure can be found in Refs.~\cite{Bellorin:2021udn,Bellorin:2021tkk}. The phase space is extended by adding the canonical pairs $(N^i,\pi_i)$, $(C^i,\bar{\mathcal{P}}_i)$, and $(\bar{C}_i,\mathcal{P}^i)$ to it. The last two pairs are the BFV ghosts. The definition of the BFV path integral is
\begin{equation}
Z =
\int \mathcal{D}V e^{iS},
\label{Z}
\end{equation}
where the measure and the action are given, respectively, by
\begin{eqnarray}
&&
\mathcal{D}V = 
\mathcal{D} g_{ij} \mathcal{D}\pi^{ij} \mathcal{D}N \mathcal{D}P_{N}\mathcal{D} N^{k}\mathcal{D}\pi_{k}
\mathcal{D} C^i \mathcal{D} \bar{\mathcal{P}}_i
\mathcal{D} \bar{C}_i \mathcal{D} \mathcal{P}^i
\times \delta(\theta_{1}) \delta(\theta_{2}) \sqrt{\det\{\theta_{p},\theta_{q}\}} \,, \;\;\;
\label{medida}
\\ &&
S=
\int dt d^{2}x \left( 
  \pi^{ij} \dot{g_{ij}} 
+ P_N \dot{N} 
+ \pi_i \dot{N}^i 
+ \bar{\mathcal{P}}_{i} \dot{C}^{i} 
+ \mathcal{P}^{i} \dot{\bar{C}}_{i} 
- \mathcal{H}_{\Psi}
\right) \,.
\label{scan} 
\end{eqnarray}
The quantum gauge-fixed Hamiltonian density is defined by
\begin{equation}
 \mathcal{H}_{\Psi} =
 \mathcal{H}_0 + \{\Psi,\Omega\}_{\text{D}} \,,
 \label{bfvhamiltonian}
\end{equation}
where $\Omega$ is the generator of the BRST symmetry, $\Psi$ is a gauge-fixing fermionic function and $\{\,,\}_{\text{D}}$ indicates Dirac brackets. According to the general formalism, to obtain $\Omega$ the algebra of spatial diffeomorphisms,
\begin{equation}
\{ \mathcal{H}_i \,, \mathcal{H}_j \} =
U_{ij}^k \mathcal{H}_k \,,
\end{equation}
is required. By using the explicit form of $U^k_{ij}$, the BRST charge becomes\footnote{In Refs.~\cite{Bellorin:2021udn,Bellorin:2021tkk} we presented the BFV quantization using a simplified notation: integrals needed for the definition of some functionals like $\Omega$ were omitted.}
\begin{equation}
 \Omega = 
 \int d^2x \left(
 \mathcal{H}_{i} C^{i} + \pi_{i} \mathcal{P}^{i} 
 - C^i \partial_i C^j \bar{\mathcal{P}}_j
 \right) \,.
\label{Omega}
\end{equation}

The delta $\delta(\theta_2)$ in the measure forces $P_N$ to be a spurious variable after the integration. Nevertheless, it is interesting to keep $P_N$ active in the formalism, specially to study the BRST transformations in the whole phase space. We promote the two deltas $\delta(\theta_1)$ and $\delta(\theta_2)$ to the quantum canonical Lagrangian by means of the integration on the Lagrange multipliers $\mathcal{A}$ and $\mathcal{B}$, respectively. Moreover, the particular algebra of constraints yields the important simplification
\begin{equation}
\sqrt{\det\{\theta_{p},\theta_{q}\}}=
\det\{\theta_{1},\theta_{2}\} 
\,.
\label{sprtbra}
\end{equation}
This determinant can also be incorporated to the canonical Lagrangian by means of a pair of fermionic ghosts, which we denote by $\eta,\bar{\eta}$. Thus, the measure of the second-class constraints becomes
\begin{equation}
\delta(\theta_{1}) \delta(\theta_{2}) \sqrt{\det\{\theta_{p},\theta_{q}\}} =	
\int\mathcal{D}\mathcal{A}\mathcal{D}\mathcal{B} 
\mathcal{D}\bar{\eta} \mathcal{D}\eta
\exp\left[
i \int dt d^dx\left(
\mathcal{A}\theta_1
+ \mathcal{B}\theta_2
+ \bar{\eta}\{\theta_1,\theta_2\}\eta
\right)
\right] \,.
\label{measurewithB}
\end{equation}

The main theorem of the BFV formalism guarantees that the path integral (\ref{Z}) is independent of the fermionic function $\Psi$. A form of the gauge-fixing condition, and hence of $\Psi$, that has been used in the quantization of the Ho\v{r}ava theory \cite{Bellorin:2021udn,Bellorin:2021tkk,Bellorin:2022qeu} is taken from the relativistic theories
\begin{eqnarray}
 &&
 \dot{N}^i - \chi^i = 0\,,
 \label{relativisticgaugephi}
 \\ &&
 \Psi =
 \bar{\mathcal{P}}_{i} N^{i} + \bar{C}_{i} \chi^{i} \,,
 \label{relativisticgaugepsi}
\end{eqnarray}
where $\chi^i$ is a free functional part of the gauge condition that must be specified to complete it. With the form of $\Omega$ given in (\ref{Omega}) and $\Psi$ in (\ref{relativisticgaugepsi}), and assuming that $\chi^i$ does not depend on the BFV ghosts, we obtain that the gauge-fixed Hamiltonian (\ref{bfvhamiltonian}) takes the generic form
\begin{equation}
\begin{split}
\mathcal{H}_{\Psi} = \, &
  \mathcal{H}_{0} 
+ \mathcal{H}_iN^{i}
+ \bar{\mathcal{P}}_i \mathcal{P}^{i}
- \bar{\mathcal{P}}_{i} \left(N^{j}\partial_{j} C^{i} +  
       N^{i} \partial_{j} C^{j} \right)
+\pi_i\chi^{i}
\\ &
+ \bar{C}_i \{\chi^{i} \,, \mathcal{H}_j\}_{\mathrm{D}} C^{j}
+ \bar{C}_i \dfrac{ \delta\chi^{i}}{\delta N^{j}} \mathcal{P}^{j} \,.
\label{genhamiltonian}
\end{split}
\end{equation}
We may take the $\chi^i$ factor that connects to the renormalizable projectable theory \cite{Barvinsky:2015kil,Bellorin:2021udn}. The perturbative variables are defined by
\begin{equation}
 g_{ij} = \delta_{ij} + h_{ij} \,,
 \quad
 \pi^{ij} = p^{ij} \,,
 \quad
 N = 1 + n \,,
\quad
N^i = n^i \,.
\end{equation}
On the rest of quantum fields, we keep the original notation, considering them of perturbative order. We denote the trace of $h_{ij}$ by $h = h_{kk}$. In $2+1$ dimensions, the required gauge-fixing condition is
\begin{equation}
	\chi^{i}= 
	\mathfrak{D}^{ij}\pi_{j} 
	- 2 \Delta\partial_jh_{ij}
	+ 2 \rho \Delta\partial_{i}h
	- 2\kappa \partial_{ijk} h_{jk} \,,
	\label{chi}
\end{equation}
where 
\begin{equation}
	\mathfrak{D}^{ij} = \delta_{ij} \Delta + \kappa \partial_{ij}  \,,
	\quad
	\Delta \equiv \partial_{kk}
	\label{D} 
\end{equation}
$\rho = \lambda ( 1 + \kappa )$, $\tilde{\rho} =  1 - 2 \lambda + 2 \kappa (1-\lambda)$ for future use, and $\kappa$ is an arbitrary constant. We use the shorthand $\partial_{ij\cdots k} = \partial_i \partial_j \cdots \partial_k$. The coefficients in $\chi^i$ has been chosen to simplify the resulting propagators of the quantum fields. Under the gauge-fixing condition (\ref{chi}), the path integral takes the form
\begin{equation}
\begin{split}
Z =& 
\int 
\mathcal{D}h_{ij} \mathcal{D}p^{ij} 
\mathcal{D}n \mathcal{D}P_N 
\mathcal{D}n^{i} \mathcal{D}\pi_{i} \mathcal{D} C^i \mathcal{D}\bar{\mathcal{P}}_i \mathcal{D}\bar{C}_i \mathcal{D}\mathcal{P}^i 
\mathcal{D}\mathcal{A} \mathcal{D}\mathcal{B}
\mathcal{D}\bar{\eta}
\mathcal{D}\eta 
\\ &
\exp\left[ i \int dt d^{2}x 
\Big(
  p^{ij}\dot{h}_{ij} 
  + P_N \dot{n}
  + \pi_{i}\dot{n}^{i}
  + \bar{\mathcal{P}}_i \dot{C}^i + \mathcal{P}^i \dot{\bar{C}}_i
  - \mathcal{H}_{0} 
  - \mathcal{H}_in^{i}
  - \bar{\mathcal{P}}_i \mathcal{P}^{i} 
  \right. 
\\ &    
  + \bar{\mathcal{P}}_{i} ( n^{j} \partial_{j} C^{i}
  + n^{i}\partial_{j} C^{j} )
  - \pi_i \mathfrak{D}^{ij} \pi_{j}
  + 2 \pi_i ( 
      \Delta\partial_jh_{ij}
	- \rho \Delta\partial_{i}h
	+ \kappa \partial_{ijk} h_{jk} )
\\ &
  \left.
  + 2 \bar{C}^{i} \left( \delta_{ij} \Delta^2 
  + \tilde{\rho} \Delta \partial_{ij} \right) C_j
  + \mathcal{A}\theta_1
  + \mathcal{B}\theta_2
  + \bar{\eta} \{ \theta_1 \,, \theta_2 \} \eta
  \Big)\right] \,.
\end{split}
\label{pathintegralguagefix}
\end{equation}
In this action, the objects $\mathcal{H}_0$, $\mathcal{H}_i$, $\theta_1$, and $\{ \theta_1 \,, \theta_2 \}$ must be expanded at the required order in perturbations.

\section{BRST symmetry}
For a theory with second-class constraints the BRST transformations are defined by means of a canonical transformation with Dirac brackets,
\begin{equation}
 \delta_\Omega \varphi = \{ \varphi \,, \Omega \}_{\text{D}} \epsilon \,,
 \label{deltaomega}
\end{equation}
where $\varphi$ stands for a generic object that depends on the canonical variables and $\epsilon$ is the fermionic parameter of the BRST transformation. The BRST charge of the Ho\v{r}ava theory is given in Eq.~(\ref{Omega}). On the canonical fields $g_{ij}$ and $\pi^{ij}$, the result of the BRST transformation is a diffeomorphism along the vector $C^i\epsilon$,\footnote{In Ref.~\cite{Barvinsky:2015kil}, using perturbative variables under the Lagrangian formalism of the projectable theory, the BRST transformation on $h_{ij}$ was identified as a diffeomorphism along $C^i$.}
\begin{eqnarray}
 &&
 \delta_\Omega g_{ij} = 
 \partial_k g_{ij} C^k \epsilon 
 + 2 g_{k(i} \partial_{j)} C^k \epsilon 
 \equiv \delta^{\text{diff}}_{C\epsilon} g_{ij} \,,
 \label{deltag}
 \\ &&
 \delta_\Omega \pi^{ij} = 
 \partial_k \pi^{ij} C^k \epsilon 
 - 2 \pi^{k(i} \partial_k C^{j)} \epsilon 
 + \pi^{ij} \partial_k C^k \epsilon 
 \equiv \delta^{\text{diff}}_{C\epsilon} \pi^{ij} \,.
\end{eqnarray}

Interestingly, the BRST transformations of other fields also involve the diffeomorphism along $C^i \epsilon$. This is the case of $N$. Let us see how this works in detail. As a consequence of dealing with Dirac brackets, there arises the factor
\begin{equation}
 \{ \theta_1 \,, \theta_2 \} =
 \frac{ \delta \theta_1 }{ \delta N } \,.
 \label{bracket}
\end{equation}
The direct computation from (\ref{deltaomega}) yields
\begin{eqnarray}
 &&
 \delta_\Omega N =
 - W \{ \theta_1 \,, \mathcal{H}_i C^i \} \epsilon \,,
 \label{deltaNprimary}
 \\ &&
 W \equiv \left( \frac{ \delta \theta_1 }{ \delta N} \right)^{-1} \,.
 \label{W}
\end{eqnarray}
We may write this transformation in a more illustrative form. $\theta_1$ is a spatial density that depends exclusively on the canonical fields $g_{ij}$, $\pi^{ij}$, and $N$. $\mathcal{H}_i$, defined in (\ref{momentumconts}), is the generator of the spatial diffeomorphisms on the canonical pair $(g_{ij},\pi^{ij})$. The complete generator that also acts on the pair $(N,P_N)$ is
\begin{equation}
 \mathcal{H}_i + \theta_2 \partial_i N \,.
 \label{completegenerator}
\end{equation}
Thus, in (\ref{deltaNprimary}) we may recover the spatial diffeomorphism on $\theta_1$ if we complete the generator (\ref{completegenerator}), compensating with an extra term,
\begin{equation}
 \delta_\Omega N =
 - W \delta^{\text{diff}}_{C\epsilon} \theta_1
 + W \{ \theta_1 \,, \theta_2 \partial_i N C^i \} \epsilon \,.
 \label{deltaNpre}
\end{equation}
Due to the functional dependence of $\theta_1$, the last bracket in this equation is simplified: $\{ \theta_1 \,, \theta_2 \partial_i N C^i \} \epsilon = \{ \theta_1 \,, \theta_2  \} \partial_i N C^i \epsilon$. Hence, Eq.~(\ref{deltaNpre}) becomes
\begin{equation}
 \delta_\Omega N =
 - W \delta^{\text{diff}}_{C\epsilon} \theta_1
 +  \partial_i N C^i \epsilon 
 = 
 - W \delta^{\text{diff}}_{C\epsilon} \theta_1
 + \delta^{\text{diff}}_{C\epsilon} N  \,.
\end{equation}

Another interesting case is the transformation of the ghost field $C^i$ itself. Its BRST transformation can be interpreted as a spatial diffeomorphism on it along the same vector $C^i\epsilon$,
\begin{equation}
\delta_\Omega C^i = 
\partial_j C^i C^j \epsilon 
\equiv \frac{1}{2} \delta^{\text{diff}}_{C\epsilon} C^i \,.
\label{diffC}
\end{equation}
This diffeomorphism of $C^i$ along itself is not zero due the Grassmann nature of $C^i$ and $\epsilon$. The last field whose BRST transformation involves a diffeomorphism is $\bar{\mathcal{P}}_i$,
\begin{equation}
\delta_\Omega \bar{\mathcal{P}}_i 
=
\left(
\partial_j \bar{\mathcal{P}}_i C^j
+ \bar{\mathcal{P}}_j \partial_i C^j
+ \bar{\mathcal{P}}_i \partial_j C^j
\right) \epsilon
+ \mathcal{H}_i \epsilon 
= 
\delta^{\text{diff}}_{C\epsilon} \bar{\mathcal{P}}_i
+ \mathcal{H}_i \epsilon \,.
\label{deltaP}
\end{equation}
The rest of transformations on the other fields are computed straightforwardly from (\ref{deltaomega}). In summary, the BRST transformations of the quantized theory are given by
\begin{equation}
\begin{array}{ll}
 \delta_\Omega g_{ij} =
 \delta^{\text{diff}}_{C\epsilon} g_{ij} \,,
 &
 \delta_\Omega \pi^{ij} =
 \delta^{\text{diff}}_{C\epsilon} \pi^{ij} \,,
 \\[1ex]
 {\displaystyle
 \delta_\Omega N =
 \delta^{\text{diff}}_{C\epsilon} N
 - W \delta^{\text{diff}}_{C\epsilon} \theta_1 } \,, 
 \qquad
 &
 \delta_\Omega P_N = 0 \,,
 \\[2ex]
 \delta_\Omega N^i = 
 \mathcal{P}^i \epsilon \,,
 &
 \delta_\Omega \pi_i = 0 \,,
 \\[1ex] {\displaystyle
 \delta_\Omega C^{i} = 
 \frac{1}{2} \delta^{\text{diff}}_{C\epsilon} C^i } \,,
 & 
 \delta_\Omega \bar{\mathcal{P}}_{i} =
 \delta^{\text{diff}}_{C\epsilon} \bar{\mathcal{P}}_i
 + \mathcal{H}_i \epsilon  \,, 
 \\[1ex]
 \delta_\Omega \mathcal{P}^i = 0 \,,
 &
 \delta_\Omega \bar{C}_i = 
 \pi_i \epsilon \,.
 \end{array}
\label{brst}
\end{equation}

Now we study the explicit operation of these transformations on the quantum action. To manage this, it is convenient to separate the quantum action in four sectors: the kinetic terms, the primary Hamiltonian $\mathcal{H}_0$, the gauge-fixing terms that come from $\{ \Psi \,, \Omega \}_{\text{D}}$, and the measure of the second-class constraints (\ref{measurewithB}). We start with the transformation of the primary Hamiltonian $\mathcal{H}_0$. One may use the fact that its integral is equivalent to the integral of the second-class constraint $\theta_1$, Eq.~(\ref{primaryhamiltonian}). By doing the functional calculus explicitly, we obtain
\begin{equation}
\begin{split}
 \delta_\Omega \int d^2x \mathcal{H}_0 \,=\, &
 \int d^2x \left(
    \frac{ \delta \theta_1 }{ \delta g_{ij} } \delta_\Omega g_{ij}
 +  \frac{ \delta \theta_1 }{ \delta \pi^{ij} } \delta_\Omega \pi^{ij}
 +  \frac{ \delta \theta_1 }{ \delta N } \delta_\Omega N 
 \right)
 \\
 = \,&  
 \int d^2x \left(
 \frac{ \delta \theta_1 }{ \delta g_{ij} } 
    \delta^{\text{diff}}_{C\epsilon} g_{ij}
 +  \frac{ \delta \theta_1 }{ \delta \pi^{ij} } 
    \delta^{\text{diff}}_{C\epsilon} \pi^{ij}
 +  \frac{ \delta \theta_1 }{ \delta N } 
    \delta^{\text{diff}}_{C\epsilon} N
 - \delta^{\text{diff}}_{C\epsilon} \theta_1
 \right)
 \\ 
 = \,&  0 \,,
\end{split}
\end{equation}
where we have used that $\theta_1$ depends only on $g_{ij}$, $\pi^{ij}$, and $N$. Note that the exact cancellation holds in the whole phase space, including the region where the constraints are not zero. This procedure can be generalized for any functional $\mathcal{F}$ that depends only on the canonical variables $g_{ij}$, $\pi^{ij}$ and $N$,
\begin{equation}
 \delta_\Omega \mathcal{F} =
  \frac{ \delta \mathcal{F} }{ \delta g_{ij} } \delta_\Omega g_{ij}
 +\frac{ \delta \mathcal{F} }{ \delta \pi^{ij} } \delta_\Omega \pi^{ij}
 +\frac{ \delta \mathcal{F} }{ \delta N } \delta_\Omega N 
 =
 \delta^{\text{diff}}_{C\epsilon} \mathcal{F}
 - \frac{\delta \mathcal{F}}{\delta N} 
   W \delta^{\text{diff}}_{C\epsilon} \theta_1 \,. 
 \label{deltaF}
\end{equation}

Next, we check the transformation of the kinetic terms. In the process we discard total time derivatives since the action of the field theory has initial and final boundaries at infinity. Thus, the transformation
\begin{equation}
 \delta_\Omega \left(
 \pi_i \dot{N}^i 
 + \mathcal{P}^{i} \dot{\bar{C}}_{i}
 \right)
 \label{firstkinetic}
\end{equation}
vanishes since it forms a total time derivative. The transformation of the other BFV ghosts can be grouped in two parts,
\begin{equation}
 \delta_\Omega \left( \bar{\mathcal{P}}_{i} \dot{C}^{i} \right)
 =
 \left( 
 \delta^{\text{diff}}_{C\epsilon} \bar{\mathcal{P}}_{i} \dot{C}^{i} 
 - \frac{1}{2} \dot{\bar{\mathcal{P}}}_{i} 
    \delta^{\text{diff}}_{C\epsilon} C^i
 \right)
 - \mathcal{H}_i \dot{C}^i \epsilon \,.
\end{equation}
By using the explicit form of the diffeomorphisms, we obtain that the terms between brackets cancel out exactly. Similarly, and using the form of $\mathcal{H}_i$ explicitly, the last term combines with the transformation
\begin{equation}
 \delta_\Omega \left( \pi^{ij} \dot{g}_{ij} \right) =
 \delta^{\text{diff}}_{C\epsilon} \pi^{ij} \dot{g}_{ij}
 - \dot{\pi}^{ij} \delta^{\text{diff}}_{C\epsilon} g_{ij}
 \label{kineticmetric}
\end{equation}
to cancel completely between them. The last kinetic term to consider is
\begin{equation}
	\delta_\Omega\left( P_N \dot{N} \right) 
	= 
	\theta_2 \partial_{t} \left( \delta_\Omega N \right) \,. 
	\label{transpdotn}
\end{equation}
This transformation does not vanish outside the constrained surface. In the following we show that this nonzero transformation can be compensated by the transformation of a Lagrange multiplier, hence the BRST invariance is kept in the full phase space.

We study the sector of the quantum Lagrangian associated with the measure of the second-class constraints, Eq.~(\ref{measurewithB}). $\mathcal{A}$ is the Lagrange multiplier of $\theta_1$. This constraint is left invariant by the BRST transformation since it is a second-class constraint. Consequently, we impose
\begin{equation}
 \delta_\Omega \mathcal{A} = 0 \,.
\end{equation}
$\mathcal{B}$ is also a multiplier of a second-class constraint, but there is still the nonzero transformation (\ref{transpdotn}) proportional to $\theta_2$. Hence we require
\begin{equation}
 \delta_\Omega\mathcal{B} = 
 - \partial_t ( \delta_\Omega N ) =
 - \partial_t \left( \delta^{\text{diff}}_{C\epsilon} N \right) 
 + \partial_t \left(
  W \delta^{\text{diff}}_{C\epsilon} \theta_1 
 \right) \,.
\end{equation}
Finally, the transformation of the ghosts $\eta,\bar{\eta}$ must be defined in such a way that the integral of the last term in (\ref{measurewithB}) remains invariant. Notice that the bracket $\{\theta_1,\theta_2\}$, given in  (\ref{bracket}), is a functional of $g_{ij}$, $\pi^{ij}$ and $N$. Thus, by using formula (\ref{deltaF}), we obtain
\begin{equation}
 \delta_\Omega \{\theta_1,\theta_2\} = 
 \delta^{\text{diff}}_{C\epsilon} 
      \left( \frac{ \delta \theta_1 }{ \delta N } \right)
 - \frac{ \delta^2 \theta_1 }{ \delta N  \delta N }
      W \delta^{\text{diff}}_{C\epsilon}  \theta_1 \,.
 \label{transbracket}
\end{equation}
We may get the combination $\bar{\eta}\{\theta_1,\theta_2\}\eta$ to transform with a diffeomorphisms along $C^i\epsilon$, hence its integral is invariant, if we define the transformation of the ghosts as follows:
\begin{eqnarray}
 \delta_\Omega \eta &=& 
 \delta^{\text{diff}}_{C\epsilon} \eta 
 + \frac{1}{2} \frac{ \delta^2 \theta_1 }{ \delta N  \delta N } W^2 \eta \,
 \delta^{\text{diff}}_{C\epsilon}  \theta_1 \,,
 \\
 \delta_\Omega \bar{\eta} &=& 
 \delta^{\text{diff}}_{C\epsilon} \bar{\eta} 
 + \frac{1}{2} \frac{ \delta^2 \theta_1 }{ \delta N  \delta N }
    W^2 \bar{\eta} \delta^{\text{diff}}_{C\epsilon}  \theta_1 \,.
\end{eqnarray}

Finally, we study the BRST transformation of the gauge-fixing terms in (\ref{pathintegralguagefix}). Since the gauge-fixing condition is imposed on perturbative variables, we study the linearized transformations. The linearized version of the BRST transformations (\ref{brst}) of the fields involved in the gauge-fixing are
\begin{equation}
\begin{array}{ll}
 \delta_\Omega h_{ij} =
 2 \partial_{(i} C_{j)} \epsilon \,,
 \hspace*{4em} &
 \delta_\Omega \pi^{ij} = 0 \,,
 \\[1ex]
 \delta_\Omega n^i = 
 \mathcal{P}^i \epsilon \,,
 &
 \delta_\Omega \pi_i = 0 \,,
 \\[1ex]
 \delta_\Omega C^{i} = 0 \,,
 & 
 \delta_\Omega \bar{\mathcal{P}}_{i} =
	\mathcal{H}_i \epsilon \,,
\\[1ex]
\delta_\Omega \mathcal{P}^i = 0 \,,
&
\delta_\Omega \bar{C}_i = 
\pi_i \epsilon \,.
\end{array}
\end{equation}
In the quantum Lagrangian (\ref{pathintegralguagefix}), the quadratic-order terms coming from the gauge-fixing are
\begin{equation}
- \mathcal{H}_in^{i}
- \bar{\mathcal{P}}_i \mathcal{P}^{i}
- \pi_i \mathfrak{D}^{ij} \pi_{j}
+ 2 \pi_i ( 
\Delta\partial_jh_{ij}
- \rho \Delta\partial_{i}h
+ \kappa \partial_{ijk} h_{jk} )
+ 2 \bar{C}^{i} \left( \delta_{ij} \Delta^2 
+ \tilde{\rho} \Delta \partial_{ij} \right) C_j  \,.
\end{equation}
Cancellations between these terms work as follows:
\begin{equation}
\begin{split}
 &
 \delta_\Omega (\mathcal{H}_i n^i ) = 
 - \delta_\Omega( \bar{\mathcal{P}}_i \mathcal{P}^i ) \,,
 \\ &
 \delta_\Omega ( \pi_i \mathfrak{D}^{ij} \pi_{j}) = 0 \,,
 \\ &
 \delta_\Omega \left[ 
 \pi_i ( 
 \Delta\partial_jh_{ij}
 - \rho \Delta\partial_{i}h
 + \kappa \partial_{ijk} h_{jk} ) 
 \right] 
 = 
 - \delta_\Omega \left[
 \bar{C}^{i} \left( \delta_{ij} \Delta^2 
 + \tilde{\rho} \Delta \partial_{ij} \right) C_j 
 \right] 
 \,.
 \end{split}
\end{equation}

A further question about the BRST transformations (\ref{brst}) is their locality. This is affected by the presence of the second-class constraints (we recall that in the BFV formalism only the second-class constraints are imposed directly). Dirac brackets are intrinsically nonlocal operations when applied to a field theory (unlike the case of a mechanical system). Therefore, nonlocal effects on the BRST transformations are to be expected. Specifically, nonlocal effects arise in the last term of $\delta_\Omega N$ in (\ref{brst}). We remark that this is not an obstruction for the consistency of the BRST transformations in the whole phase space, as we have proven. Moreover, by evaluating the transformation of $N$ on the surface where the second-class constraints are satisfied, hence $\theta_1 = 0$ and $\delta^{\text{diff}}_{C\epsilon} \theta_1 = 0$, it becomes a pure diffeomorphism on $N$,
\begin{equation}
 \delta_\Omega N = \delta^{\text{diff}}_{C\epsilon} N \,.
\end{equation}
This transformation is a local operation. The rest of the transformations in (\ref{brst}) are local. Therefore, we find that the BRST symmetry transformations are strictly local operations when the theory is restricted to the surface where the second-class constraints are satisfied. The Hamiltonian, with the gauge-fixing condition, is also local.

\section{Nilpotence of the BRST transformations}

It is illustrative to see explicitly how the nilpotence of the BRST transformations is based largely on the diffeomorphism along $C^i\epsilon$. We can study this on nonperturbative variables and on the whole phase space.

We introduce the BRST operator $s$, such that the BRST transformation of a field $\phi$ is $\delta_\Omega \phi = \epsilon s \phi$. Let us start with the case of the spatial metric,
\begin{equation}
 s g_{ij} = 
 - \partial_k g_{ij} C^k - 2 g_{k(i} \partial_{j)} C^k \,.
\end{equation}
We compute the square action of the BRST transformation in the form $\delta_\Omega (sg_{ij})$ \cite{Weinberg:1996kr}. By operating this transformation on $sg_{ij}$ and substituting (\ref{deltag}) and (\ref{diffC}) into it, we get that it cancels completely. The square transformation $\delta_\Omega (s\pi^{ij})$ is a computation parallel to this one. From the transformation of $N$ in (\ref{brst}) we obtain
\begin{eqnarray}
 &&
 s N =
 - \partial_i N C^i 
 + W \delta^{\text{diff}}_C \theta_1 \,,
 \\ &&
 \delta^{\text{diff}}_C \theta_1 \equiv
 \partial_i \theta_1 C^i + \theta_1 \partial_i C^i  \,.
\end{eqnarray}
Note that the diffeomorphism $\delta^{\text{diff}}_C \theta_1$ is of Grassmann nature and does not involve the parameter $\epsilon$. When computing the square $\delta_\Omega(sN)$, a required term is $\delta_\Omega W \delta^{\text{diff}}_C \theta_1$.  $W$, given in (\ref{W}), depends only on $g_{ij}$, $\pi^{ij}$ and $N$; hence formula (\ref{deltaF}) applies:
\begin{equation}
\delta_\Omega W \delta^{\text{diff}}_C \theta_1 = 
\left(  
\delta^{\text{diff}}_{C\epsilon} W 
- \frac{ \delta W }{ \delta N } W \delta^{\text{diff}}_{C\epsilon} \theta_1
\right)
\delta^{\text{diff}}_C \theta_1
= 
\delta^{\text{diff}}_{C\epsilon} W \delta^{\text{diff}}_C \theta_1 \,. 
\label{deltaW}
\end{equation}
In the last equality we have used $ \delta^{\text{diff}}_{C\epsilon} \theta_1 \times \delta^{\text{diff}}_C \theta_1 = 0$, due to the Grassmann nature of $C^i$. Transformation (\ref{deltaW}) requires a diffeomorphism on $W$, which is a scalar density of weight $-1$. Another term that is required is
\begin{equation}
 W \delta_\Omega ( \delta^{\text{diff}}_C \theta_1 ) =
 - W \left ( \partial_i \theta_1 \delta_\Omega C^i 
           + \theta_1 \partial_i \delta_\Omega C^i \right) \,,
 \label{deltadeltatheta}
\end{equation}
where we have used $\delta_\Omega \theta_1 = 0$ since $\theta_1$ is a second-class constraint. After using (\ref{deltaW}) and (\ref{deltadeltatheta}), the vanishing of the square transformation $\delta_\Omega ( sN )$ becomes a straightforward computation. The vanishing of $\delta_\Omega ( s C^i )$ is a direct result from (\ref{diffC}). 
The transformation $s \bar{\mathcal{P}}_i$ is extracted from (\ref{deltaP}). When computing the square $\delta_\Omega ( s \bar{\mathcal{P}}_i )$, a useful fact is $\delta_\Omega \mathcal{H}_i = \delta^{\text{diff}}_{C\epsilon} \mathcal{H}_i$, since $\mathcal{H}_i$ is a functional only of $g_{ij}$ and $\pi^{ij}$ and the formula (\ref{deltaF}) can be applied for it. In addition, the transformation of the terms linear in $\bar{\mathcal{P}}_i$ in (\ref{deltaP}) generates terms linear in $\mathcal{H}_i$; they form another $\delta^{\text{diff}}_{C\epsilon} \mathcal{H}_i$ factor that cancels the previous one. The rest is a straightforward computation to arrive at the vanishing of $\delta_\Omega ( s \bar{\mathcal{P}}_i )$. The remaining fields in (\ref{brst}) have an obvious nilpotent action of the BRST transformation.

\section{Unitarity of the nonprojectable case}

The unitarity of the $S$ matrix holds if the path integral can be formulated as an integral over physical or reduced phase space with measure 1 (that is, the measure only contains the differentials of the independent physical fields). Previously we discussed that the $2+1$ theory must describe the physics of a scalar mode (the so-called extra scalar mode of the Ho\v{r}ava theory). We first check that the global balance in the quantum theory matches with this: in (\ref{Z}) the bosonic canonical fields $\{ g_{ij},\pi^{ij},N,P_N,N^i,\pi_i \}$ sum up 12 independent degrees of freedom. The BFV ghosts $\{ C^i,\bar{\mathcal{P}}_i,\bar{C}_i,\mathcal{P}^i \}$ are introduced to reduce the dynamics on the phase space, and they sum up 8 degrees of freedom. There are two Dirac deltas for the second-class constraints; hence there remains 2 degrees of freedom in the physical phase space. This is the quantum scalar mode.

The reduction of the phase space requires us to solve the second-class constraints. Below we shall see that the measure leads naturally to solve the constraint $\theta_1$ for $N$; hence this is a requisite for unitarity. We discuss first the existence of the solution. By considering the complete $2+1$ theory with the terms of order $z=2$, we have that the $\theta_1 = 0$ constraint (\ref{theta1}) can be posed as a fourth-order nonlinear differential equation for $N$:
\begin{eqnarray}
&& 
2\alpha_{6} N^{-1} \nabla^{i}\nabla^{j} (N\nabla_{j}a_{i})
+ 2\alpha_{7} N^{-1} \nabla^{2}(N\nabla_{i}a^{i})
+ \alpha_{4} \left( N^{-1} \nabla^{2}(Na^{2}) 
- 2 \nabla_{i}( a^{i} \nabla_{j} a^{j} ) 
- a^{2}\nabla_{i}a^{i}
\right)
\nonumber \\ && 
+ 3 \alpha_{6} \nabla^{i}a^{j}\nabla_{i}a_{j}
+ \alpha_{7}(\nabla_{i}a^{i})^{2}  
- 4 \alpha_{2} \nabla_{i}( a^{2} a^{i} )
+ \alpha_{5} \left( R\nabla_{i}a^{i} 
+ N^{-1} \nabla^{2} (NR) \right)
\nonumber\\ &&
- \alpha_{3} \left( 2 \nabla_{i}( R a^{i} )
+ R a^{2} \right)
+ \alpha \left( 2 \nabla_{i} a^{i}
+ a^{2}  \right)
- 3 \alpha_{2} a^{4}
\nonumber \\ &&
= 
- \alpha_{1} R^{2}
+ \beta R
- g^{-1} \left( \pi^{ij}\pi_{ij}
+ \bar{\sigma} \pi^{2} \right)
\,.
\label{eqNnonperturba}
\end{eqnarray}
Its highest-order derivative comes from the first two terms, resulting in
\begin{equation}
	2 ( \alpha_6 + \alpha_7 ) g^{ij} g^{kl} \partial_{ijkl} \ln N \,.
	\label{higherorder}
\end{equation}
Since $g_{ij}$ is a Riemannian metric, this term is an elliptic operator acting on $\ln N$. The linearized version of Eq.~(\ref{eqNnonperturba}) is a fourth-order elliptic equation,
\begin{equation}
	2 ( \alpha_6 + \alpha_7 ) \Delta^2 n
	+ 2 \alpha \Delta n
	= 
	\left( - \alpha_{5} \Delta + \beta \right) 
	\left( - \Delta h + \partial_{ij} h_{ij} \right)
	\label{diffeqN}
	\,.
\end{equation}
To hold this feature, we require 
\begin{equation}
\alpha_6 + \alpha_7 \neq 0   \,.
\label{condalfa6alfa7}
\end{equation}
To talk about the existence of the solution of Eq.~(\ref{diffeqN}), we adopt the exposition of Ref.~\cite{Bers}, which uses the Hilbert space approach on linear elliptic equations of higher order. According to the theorems collected in this book \cite{Bers}, if a Dirichlet problem is posed with Eq.~(\ref{diffeqN}), the weak solution exists if the associate homogeneous equation $2 ( \alpha_6 + \alpha_7 ) \Delta^2 n
+ 2 \alpha \Delta n
= 0$ has $n= 0$ as its only solution.

In the high-energy regime, the low-order derivatives in Eq.~(\ref{diffeqN}) can be neglected, such that the resulting equation has only the squared Laplacian operator on $n$,
\begin{equation}
2 ( \alpha_6 + \alpha_7 ) \Delta^2 n
= 
- \alpha_{5} \Delta 
\left( - \Delta h + \partial_{ij} h_{ij} \right)
\label{diffeqNUV}
\,.
\end{equation}	
In this case the solution can be given in a closed way by means of the convolution with the fundamental solution of $\Delta^2$ in two dimensions, which is
\begin{equation}
 \Delta^2 K(r) = \delta(r) \,,
 \qquad
 K(r) = \frac{1}{8\pi} r^2 \ln r \,,
\end{equation}
where $r \equiv \sqrt{ x_1^2 + x_2^2 }$. For $d=3$ and other dimensions, the fundamental solution of the operator $\Delta^z$ can be found in Ref.~\cite{John}. Moreover, in Ref.~\cite{John} the fundamental solution for a more general elliptic operator, having nonconstant coefficients or low-order derivatives as (\ref{diffeqN}), is presented in terms of plane waves (see also \cite{Bers}).

Further evidence for the existence of the solution of (\ref{eqNnonperturba}) comes from a different limit: the large-distance limit on nonperturbative variables. Under this approximation, discarding all spatial derivatives of fourth order, Eq.~(\ref{eqNnonperturba}) becomes a linear elliptic equation of second-order for $\sqrt{N}$ (see \cite{Bellorin:2011ff}),
\begin{equation}
 4 \alpha \nabla^2 \sqrt{N} 
 - \left( 
 \beta R - g^{-1} ( \pi^{ij} \pi_{ij} + \bar{\sigma} \pi^2 ) \right)
 \sqrt{N}
 = 0 
\label{eqsqrtN}
\end{equation}
(whenever $\alpha \neq 0$). Summarizing, the two limiting cases given in Eqs.~(\ref{diffeqNUV}) and (\ref{eqsqrtN}) have solutions, for the linearized equation (\ref{diffeqN}) there is a theory under which the weak solution can be addressed, whereas other approaches are available, and the complete Eq.~(\ref{eqNnonperturba}) remains as a nonlinear equation with a highest-order operator that is elliptic.

To make the reduction of the path integral, we return to an undefined gauge-fixing function $\chi^i$; condition (\ref{chi}) is not imposed. The path integral is taken in its form (\ref{Z}), with the Hamiltonian $\mathcal{H}_\Psi$ given in (\ref{genhamiltonian}). We show that the unitarity of the $S$ matrix naturally leads to the necessity of the solution of $\theta_1$ in terms of $N$: the delta $\delta (\theta_2)$ sets $P_N = 0$, eliminating the kinetic term $P_N \dot{N}$. According to (\ref{bracket}), the measure associated with the second-class constraints becomes
\begin{equation}
 \delta( \theta_1 ) \det\left( \frac{ \delta \theta_1 }{ \delta N } \right)
 = \delta( N - \hat{N} ) \,,
 \label{deltaN}
\end{equation}
where we have posed the constraint $\theta_1 = 0$ as an equation for $N$, and we have denoted by $\hat{N}$ its solution, assuming that it exists at arbitrary order or at nonperturbative level. Thus, the measure of the second-class constraints naturally leads to fixing $N$ after the reduction of the phase space. The path integral takes the form
\begin{equation}
\begin{split}
Z = & \int \mathcal{D} g_{ij} \mathcal{D}\pi^{ij} \mathcal{D} N^{k}\mathcal{D}\pi_{k}
\mathcal{D} C^i \mathcal{D} \bar{\mathcal{P}}_i
\mathcal{D} \bar{C}_i \mathcal{D} \mathcal{P}^i
\\ &
\times \exp \left[ i \int dt d^{2}x \left( 
\pi^{ij} \dot{g_{ij}} 
+ \pi_i \dot{N}^i 
+ \bar{\mathcal{P}}_{i} \dot{C}^{i} 
+ \mathcal{P}^{i} \dot{\bar{C}}_{i} 
- \mathcal{H}_{\Psi} |_{N = \hat{N}}
\right) \right] \,.
\end{split}
\end{equation}
$\mathcal{H}_\Psi$ depends functionally on $N$ only through $\mathcal{H}_0$ and the gauge-fixing condition $\chi^i$. The second-class constraints have been eliminated, but the phase space still contains several unphysical variables.

To arrive at the final unitary form, we apply the procedure developed in Refs.~\cite{Faddeev:1969su,Fradkin:1977hw}. The first step is to integrate the BFV ghosts. Suppose that $\Phi^i(g_{ij}, \pi^{ij}) = 0$ is a gauge-fixing condition of interest. To adapt the BFV quantization to this gauge we set
\begin{equation}
 \chi^i = \frac{1}{\varepsilon} \Phi^i(g_{ij},\pi^{ij}) \,,
 \label{chiunitary}
\end{equation} 
where $\varepsilon$ is an arbitrary numerical parameter. The BFV theorem ensures that the resulting path integral is independent of $\varepsilon$. Hence, by taking at the end the limit $\varepsilon \rightarrow 0$ we recover the desired gauge-fixing condition $\Phi^i = 0$. Simultaneously, the following rescaling on $\pi_i$ and $\bar{C}_i$ is done,
\begin{equation}
 \pi_i \rightarrow \varepsilon \pi_i \,,
 \quad
 \bar{C}_i \rightarrow \varepsilon \bar{C}_i \,.
 \label{rescaling}
\end{equation}
Since $\bar{C}_i$ is a Grassmann variable, the Jacobian of this rescaling is one. The Hamiltonian (\ref{genhamiltonian}) takes the form
\begin{equation}
\mathcal{H}_{\Psi}|_{N=\hat{N}} = 
\mathcal{H}_{0}|_{N=\hat{N}} 
+ \mathcal{H}_iN^{i}
+ \bar{\mathcal{P}}_i \mathcal{P}^{i}
- \bar{\mathcal{P}}_{i} \left(N^{j}\partial_{j} C^{i} +  
N^{i} \partial_{j} C^{j} \right)
+\pi_i\Phi^{i}
+ \bar{C}_i \{\Phi^i \,, \mathcal{H}_j\} C^{j} \,.
\label{Hamiltoniancanonnicalgauge}
\end{equation}
$\mathcal{H}_{0}|_{N=\hat{N}}$ depends exclusively on the canonical pair $(g_{ij},\pi^{ij})$. Note that the last term of this Hamiltonian has a Poisson bracket indicated, instead of a Dirac bracket. This is consistent with the fact that we have solved the second-class constraints. The path integral takes the form
\begin{equation}
\begin{split}
Z = & \int \mathcal{D} g_{ij} \mathcal{D}\pi^{ij} \mathcal{D} N^{k}\mathcal{D}\pi_{k}
\mathcal{D} C^i \mathcal{D} \bar{\mathcal{P}}_i
\mathcal{D} \bar{C}_i \mathcal{D} \mathcal{P}^i
\\ &
\times \exp \left[ i \int dt d^{2}x \left( 
\pi^{ij} \dot{g_{ij}} 
+ \varepsilon \pi_i \dot{N}^i 
+ \bar{\mathcal{P}}_{i} \dot{C}^{i} 
+ \varepsilon \mathcal{P}^{i} \dot{\bar{C}}_{i} 
- \mathcal{H}_{\Psi} |_{N = \hat{N}}
\right) \right] \,.
\end{split}
\end{equation}
Now we may take the limit $\varepsilon \rightarrow 0$, obtaining
\begin{equation}
\begin{split}
Z = & \int \mathcal{D} g_{ij} \mathcal{D}\pi^{ij} \mathcal{D} N^{k}\mathcal{D}\pi_{k}
\mathcal{D} C^i \mathcal{D} \bar{\mathcal{P}}_i
\mathcal{D} \bar{C}_i \mathcal{D} \mathcal{P}^i 
\\ &
\times \exp \left[ i \int dt d^{2}x \left( 
  \pi^{ij} \dot{g_{ij}} 
- \mathcal{H}_{0}|_{N=\hat{N}} 
- \mathcal{H}_iN^{i}
- \pi_i \Phi^i
\right.\right.\\ & \left. \left.
- \bar{\mathcal{P}}_{i} \left( 
  \mathcal{P}^{i}
- \dot{C}^{i} 
- N^{j}\partial_{j} C^{i} - N^{i} \partial_{j} C^{j}
\right) 
- \bar{C}_i \{\Phi^i \,, \mathcal{H}_j\} C^{j} 
\right) \right] \,.
\end{split}
\end{equation}
A shift on the $\mathcal{P}^i$ field with unit Jacobian leads us to the quadratic form $-\bar{\mathcal{P}}_i \mathcal{P}^i$; hence the integration on these momenta can be done with no consequences on the path integral. Next, the integration on $N^i$, $\pi_i$, $\bar{C}_i$, and $C^i$ leads to the form of the path integral,
\begin{equation}
 Z = 
 \int \mathcal{D} g_{ij} \mathcal{D}\pi^{ij} 
 \delta( \mathcal{H}_i ) \delta( \Phi^i )  
 \det \{\Phi^i \,, \mathcal{H}_j\}
 \exp \left[ i \int dt d^{2}x \left( 
 \pi^{ij} \dot{g_{ij}}
 - \mathcal{H}_{0}|_{N=\hat{N}} 
 \right) \right] \,.
 \label{pathintegralfaddeev}
\end{equation}
This is the Faddeev formula for the path integral of a system with first-class constraints only \cite{Faddeev:1969su}. The first-class constraint is $\mathcal{H}_i(g_{ij},\pi^{ij}) = 0$, and the canonical gauge-fixing condition is $\Phi^i(g_{ij},\pi^{ij}) = 0$. The last step is to show that this path integral can be formulated strictly as an integral over the physical phase space with measure 1, yielding a unitary $S$ matrix. We apply Faddeev's procedure to achieve this \cite{Faddeev:1969su}. It consists of making a canonical transformation on the phase space with the aim of identifying a part of the new canonical variables directly with the unphysical degrees of freedom and the rest with the physical ones. The consistent formulation of a system with first-class constraints requires that the gauge-fixing condition $\Phi^i$ satisfies
\begin{eqnarray}
&&
\det \{ \Phi^i \,, \mathcal{H}_j \} \neq 0 \,,
\label{phinotcommute}
\\ &&
\{ \Phi^i \,, \Phi^j \}=0 \,. 
\label{phicommutes}
\end{eqnarray} 
We perform a canonical transformation on the coordinates $(g_{ij},\pi^{ij})$, where the new coordinates are labeled by the two sets of canonical fields: $(q_i,p^i)$ and $(q^{*},p^{*})$.\footnote{In arbitrary spatial dimensions $d$, the splitting is $(q_i,p^i)$ and $(q^{*A},p^{*A})$, with $i=1,\dots,d$ and $A=1,\ldots,\frac{1}{2} d(d-1)$.} The canonical transformation preserves the kinetic terms in (\ref{pathintegralfaddeev}) and has unit Jacobian since it is done on bosonic variables. By virtue of condition (\ref{phicommutes}), we can use the canonical transformation to make the identification
\begin{equation}
 \Phi^i(g_{kl},\pi^{kl}) = p^i \,.
 \label{gaugeidentification}
\end{equation}
With this setting the condition (\ref{phinotcommute}) takes the form
\begin{equation}
 \det \left( \frac{ \partial \mathcal{H}_j }{ \partial q_i} \right) \neq 0 \,.
\end{equation}
This implies that the constraint $\mathcal{H}_j = 0$ can be solved for the $q_i$ variables. On the basis of (\ref{gaugeidentification}) and the solution of $\mathcal{H}_j = 0$, the physical phase space is defined by the equations,
\begin{equation}
 p^i = 0 \,,
 \quad
 q_i = q_i(q^*,p^{*}) \,,
 \label{solutionH} 
\end{equation}
such that $(q^*,p^{*})$ are free coordinates on the physical phase space. We can transform the factors of the measure in (\ref{pathintegralfaddeev}) as follows:
\begin{equation}
 \delta( \mathcal{H}_i ) \det \{\Phi^i \,, \mathcal{H}_j\} =
 \delta( \mathcal{H}_i ) \det \left( \frac{ \partial \mathcal{H}_j }{ \partial q_i } \right) =
 \delta(q_i - q_i(q^*,p^{*}) ) \,,
\end{equation}
where $q_i(q^*,p^*)$ identifies the solution of $\mathcal{H}_i = 0$ (\ref{solutionH}). Therefore, the integration on $q_i$ sets $q_i = q_i(q^*,p^*)$ and the integration on $p^i$ sets $p^i = 0$. There are no more factors remaining in the measure. The path integral (\ref{pathintegralfaddeev}) becomes
\begin{equation}
 Z = 
 \int \mathcal{D} q^* \mathcal{D} p^{*} 
 \exp \left[ i \int dt d^{2}x \left( 
 p^{*} \dot{ q^{*} }
 - \mathcal{H}_{0}|_{\Gamma^*} 
 \right) \right] \,.
\end{equation}
where $\Gamma^*$ is the subset of the phase space defined by
\begin{equation}
 N = \hat{N} \,,
 \quad
 q_i = q_i(q^*,p^{*}) \,,
 \quad
 p^i = 0 \,.
\end{equation}
$\Gamma^*$ is the physical phase space, since any unphysical variable has been eliminated to define it. In the present theory, $(q^*,p^*)$ are the coordinates of the physical scalar mode, and the physical Hamiltonian $\mathcal{H}_{0}|_{\Gamma^*}$ depends exclusively on them.


\section{Projectable case}
The projectable version of the Ho\v{r}ava theory is defined by the condition that the lapse function is a function only of time, $N = N(t)$. This condition is preserved by the gauge-symmetry group of the theory, the foliation-preserving diffeomorphisms (see Eq.~(\ref{deltadiffN})). Therefore, this case constitutes an independent formulation of the theory. In this case $N(t)$ is a pure gauge degree of freedom; we may use the symmetry transformation $\delta t = f(t)$ to set $N=1$, a condition that we assume throughout this section. The Lagrangian of the projectable case is given by (\ref{classicalaction}), where the potential $\mathcal{V}(g_{ij})$ depends only on the spatial metric. In $2+1$ dimensions, excluding the term linear in the spatial Ricci scalar since it leads to a topological invariant, the $z=2$ potential is given by
\begin{eqnarray}
\mathcal{V}=\alpha_0 R^2.
\end{eqnarray}

In the classical canonical formulation \cite{Kobakhidze:2009zr}, the canonical pair is $(g_{ij},\pi^{ij})$. The central difference between the dynamics of the nonprojectable and projectable case is that the latter has only first-class constraints, which is given by the momentum constraint (\ref{momentumconts}). An analog of the local ``Hamiltonian'' constraint is absent. Instead, here there arises the global (integrated) constraint,
\begin{equation}
    \int\,d^2x\mathcal{H}=0,\qquad \mathcal{H}=\frac{1}{\sqrt{g}}\left(\pi^{ij}\pi_{ij}
    +\frac{\lambda}{1-d\lambda}\pi^2\right)
    +\sqrt{g}\mathcal{V} \,.
\end{equation}
The primary Hamiltonian density can be determined by
\begin{equation}
    H_0= \int d^2x\mathcal{H}_0= \int d^2x\mathcal{H}.
\end{equation}
Thus, the primary Hamiltonian density is equivalent to $\mathcal{H}$. The classical functional variables are $\{ g_{ij},\pi^{ij} \}$, which sum six. The constraint $\mathcal{H}_i$ has $2$ degrees and there are $2$ more degrees in the spatial diffeomorphisms. The resulting physical phase space has dimension two, the same as the nonprojectable case.

We show a summary of the BFV quantization of the projectable theory. Details can be found in Ref.~\cite{Bellorin:2021udn}. The BFV path integral is
\begin{equation}
\begin{split}
Z =&
\int \mathcal{D} g_{ij} \mathcal{D}\pi^{ij} \mathcal{D}N^{k} \mathcal{D}\pi_{k} \mathcal{D} C^i \mathcal{D}\bar{\mathcal{P}}_i \mathcal{D} \mathcal{P}^i \mathcal{D} \bar{C}_i
\\ & 
\exp\left[ i \int \,dt \,d^{2}x 
\Big(\pi^{ij}\dot{g_{ij}} +\pi_{k}\dot{N}^{k}
+\bar{\mathcal{P}}_i \dot{C}^i +  \mathcal{P}^i \dot{\bar{C}}_i
- \mathcal{H}_\Psi\Big)  \right]
\,.
\end{split}
\label{bfvfinalprojectable}
\end{equation}
The BRST generator $\Omega$ has the same form of the nonprojectable case, Eq.~(\ref{Omega}). By adopting the fermionic gauge-fixing function (\ref{relativisticgaugepsi}), we obtain the gauge-fixed Hamiltonian
\begin{equation}
\mathcal{H}_{\Psi} = 
\mathcal{H} 
+\mathcal{H}_kN^{k}
+ \bar{\mathcal{P}}_k \mathcal{P}^{k}
- \bar{\mathcal{P}}_{i} \left(N^{j}\partial_{j}C^{i} + N^{i}\partial_{j}C^{j}\right)
+ \pi_k\chi^{k}
+ \bar{C}_i \{\chi^{i} \,, \mathcal{H}_k\}  C^{k}
+ \bar{C}_i \dfrac{ \delta\chi^{i}}{\delta N^{l}} \mathcal{P}^{l} \,,
\end{equation}
where the gauge-fixing condition is independent of the BFV ghost. 

Since this is a theory with first-class constraints only, the BRST symmetry transformations with parameter $\epsilon$ are defined by
$\delta_\Omega \varphi = \{ \varphi \,, \Omega \} \epsilon$. The transformations result to be the same as the nonprojectable case (\ref{brst}), if the pair $N,P_N$ is excluded. Thus, we have again that the spatial diffeomorphism along the vector $C^i\epsilon$ plays a prominent role. The primary Hamiltonian is BRST invariant:
\begin{equation}
\delta_\Omega H_0 =
\int d^2x \left(
\frac{\delta\mathcal{H}}{\delta g_{ij}}\delta_\Omega g_{ij}
+\frac{\delta\mathcal{H}}{\delta \pi^{ij}}\delta_\Omega \pi^{ij}\right)
=
\int d^2x\: \delta^{\text{diff}}_{C\epsilon} \mathcal{H}
=0 \,.
\end{equation}
Since the BRST transformations are the same, the invariance of the kinetic terms of the action follows the same steps from (\ref{firstkinetic}) to (\ref{kineticmetric}). If we adopt the gauge-fixing form (\ref{chi}), then the proof of the invariance of the terms associated with the gauge fixing is parallel to the nonprojectable case. This completes the invariance of the action.

To prove the unitarity of the $S$ matrix, we manage the integration on the ghost fields. We assume that the gauge-fixing condition has the form (\ref{chiunitary}) and we rescale the fields $\pi^i$ and $\bar{C}_i$ as in (\ref{rescaling}). We take the limit $\varepsilon\rightarrow 0$ and then we integrate on $\mathcal{P}^i, \bar{\mathcal{P}}_i$. The path integral takes the form
\begin{equation}
\begin{split}
Z =&
\int \mathcal{D} g_{ij} \mathcal{D}\pi^{ij} \mathcal{D}N^{k} \mathcal{D}\pi_{k} \mathcal{D} C^i \mathcal{D} \bar{C}_i
\\ & 
\times \exp\left[ i \int \,dt \,d^{2}x 
\Big(\pi^{ij}\dot{g_{ij}} 
-\mathcal{H} -N^k\mathcal{H}_k-\pi_k\Phi^k-\bar{C}_i\{\Phi^i,\mathcal{H}_k\}C^k\Big)  \right]
\,.
\\
 =&
\int \mathcal{D} g_{ij} \mathcal{D}\pi^{ij} \delta(\Phi^k)\delta(\mathcal{H}_{i})\det\{\Phi^j,\mathcal{H}_j\}
\exp\left[ i \int \,dt \,d^{2}x 
\Big(\pi^{ij}\dot{g_{ij}} 
-\mathcal{H} \Big)  \right]
\,.
\end{split}
\end{equation}
We have arrived at the Faddeev form of the $S$ matrix \cite{Faddeev:1969su}. The action has a canonical form on the variables $(g_{ij},\pi^{ij})$; $\mathcal{H}$ is a functional of $(g_{ij},\pi^{ij})$ from the beginning. By repeating the same steps of section 5, the unitary form of the $S$ matrix can be proven, describing the quantum dynamics of a scalar mode.

\section*{Conclusions}
We have succeed in finding explicit expressions for the BRST symmetry transformations of the Ho\v{r}ava theory, both in its nonprojectable and projectable versions, under the BFV formalism. These expressions have allowed us to prove the BRST invariance of the quantum action explicitly. The consistency of the BRST symmetry is a fundamental aspect for the quantization of the theory. This may be an input for a future proof of the renormalization of the nonprojectable case. We find a very useful result in the fact that these transformations can be cast in terms of a diffeomorphism along the ghost vector field $C^i$, including the transformation of the ghost itself and its conjugate momentum. The transformation of various objects that are tensors or tensor densities can be managed in terms of the spatial diffeomorphism. We have done the analysis in the original nonperturbative variables and in the whole phase space, including the region where the second-class constraints are not satisfied. Regarding this, the BRST transformations are strictly local when evaluated on the constrained surface, where the quantum theory is defined. Outside this surface, there are nonlocal contributions.

We have proved that, if we assume that the solution of the $\theta_1 = 0$ constraint in terms of the lapse function $N$ exists, then the unitarity of the theory holds, including the case when the theory is quantized under the gauge required for renormalization, thanks to the independence of the BFV formalism on the gauge chosen. The necessity of the solution for $N$ arises naturally in the reduction to the physical degrees of freedom, since the quantum measure leads to solve the constraint $\theta_1 = 0$ for $N$, a condition that is not evident at the level of the classical theory. This constraint is a nonlinear equation for $N$. We have appealed to its linearized version to arrive at a linear elliptic equation of fourth order (in the $2+1$ case). According to the results in the literature, existence in a general higher-order linear equation has been proven under certain conditions. In the high-energy limit of the linearized theory, the solution can be given in a closed way by the convolution with the fundamental solution, which is known. Another limit we have presented is the large-distance limit. In this case, $\theta_1 = 0$ becomes a linear elliptic equation of second order, which is a well-known case, and it is of nonperturbative character. These evidences lead us to believe that the solution for $N$ of the $\theta_1 = 0$ constraint exists at any order.


\section*{Acknowledgments}
C.B.~is partially supported by Grant No. CONICYT PFCHA/DOCTORADO BECAS CHILE /2019 -- 21190960. C.B.~is partially supported as a graduate student in the ``Doctorado en F\'isica Menci\'on F\'isica-Matem\'atica" Ph.D. program at the Universidad de Antofagasta. 



\end{document}